\documentclass[manuscript,nonacm]{acmart}

\AtBeginDocument{%
  }

\acmConference[CHI '26]{The 2026 CHI Conference on Human Factors in Computing Systems}{April 13 -- 17, 2026}{Barcelona, Spain}

\usepackage{multirow}
\usepackage{array}
\usepackage{makecell}
\usepackage[table]{xcolor}
\usepackage{graphicx}
\usepackage{float}

\begin{document}

\title[Mapping data literacy trajectories in K--12 education]{Mapping data literacy trajectories in K--12 education}

\author{Robert Whyte}
\email{bobby.whyte@raspberrypi.org}
\orcid{0000-0002-1214-1957}
\affiliation{%
  \institution{Raspberry Pi Foundation}
  \city{Cambridge}
  \country{United Kingdom}
}
\author{Manni Cheung}
\email{manni.cheung@raspberrypi.org}
\orcid{0009-0000-8290-6380}
\affiliation{
  \institution{Raspberry Pi Foundation}
  \city{Cambridge}
  \country{United Kingdom}
}

\author{Katharine Childs}
\email{katharine@raspberrypi.org}
\orcid{0009-0001-7185-6784}
\affiliation{
  \institution{Raspberry Pi Foundation}
  \city{Cambridge}
  \country{United Kingdom}
}

\author{Jane Waite}
\email{jane.waite@raspberrypi.org}
\orcid{0000-0002-0270-2124}
\affiliation{
  \institution{Raspberry Pi Foundation}
  \city{Cambridge}
  \country{United Kingdom}
}

\author{Sue Sentance}
\email{ss2600@cam.ac.uk}
\orcid{0000-0002-0259-7408}
\affiliation{
  \institution{University of Cambridge}
  \city{Cambridge}
  \country{United Kingdom}
}

\renewcommand{\shortauthors}{Whyte et al.}

\begin{abstract}
Data literacy skills are fundamental in computer science education. However, understanding how data-driven systems work represents a paradigm shift from traditional rule-based programming. We conducted a systematic literature review of 84 studies to understand K--12 learners' engagement with data across disciplines and contexts. We propose the data paradigms framework that categorises learning activities along two dimensions: (i) logic (knowledge-based or data-driven systems), and (ii) explainability (transparent or opaque models). We further apply the notion of learning trajectories to visualize the pathways learners follow across these distinct paradigms. We detail four distinct trajectories as a provocation for researchers and educators to reflect on how the notion of data literacy varies depending on the learning context. We suggest these trajectories could be useful to those concerned with the design of data literacy learning environments within and beyond CS education.

\end{abstract}



\maketitle

\section{Introduction}
\label{sect:introduction}
Data literacy is increasingly positioned as a foundational competency in a data-driven society, spanning domains such as visualization, cognitive science, artificial intelligence (AI) and education \cite{CHI_EA_2026_DataLiteracy_Workshop}. In computer science (CS) education, the rise of data-driven technologies (e.g., AI and machine learning) has led to calls for CS to expand to incorporate relevant data literacy skills \cite{LongMagerko}. However, traditional programming instruction, which focuses on rule-based logic, is insufficient for helping students master the data-driven nature of these emerging technologies \cite{tedre2021ct}. 
Likewise, what data literacy skills are needed is uncertain. Across curricula, learners may be introduced to concepts and skills relating to coding, database work, statistical analysis, and data ethics, yet the emphasis placed on these skills and how they relate to one another can differ by context. \citeauthor{CruickshankWholeElephant} argues that data science’s cross-disciplinary nature can produce a \textit{``fractured perspective''} \cite[p.~248]{CruickshankWholeElephant} in which connections between contributing domains (e.g., statistics, computer science, mathematics) remain underdeveloped or invisible. To address this, they propose a more integrated framework that makes core competencies, such as domain understanding, problem formulation, and data management, explicit across the data lifecycle, and clarifies how these competencies align across disciplines. On the other hand, \citet{Olari2024Concepts,Olari2024Practices} outline a set of concepts and practices they argue are fundamental to a `data-centric' CS education. These framings suggest that \textit{`mapping the field'} is an ongoing project that warrants an interdisciplinary view and that further theorisation is needed of \textit{what} data literacy looks like in CS education and \textit{how} instructional experiences should be structured into meaningful progressions. This position paper is therefore guided by the question: \textit{How can we map the learning experiences taken by K--12 students across interdisciplinary data literacy activities through a common framework?}

\section{Theoretical framework}
\label{ref:theoreticalFramework}
Prior research on AI literacy in K--12 settings has often included learning about ML, yet in practice this relies on students’ capability to reason with data, for example, by trying out different examples and labels and observing how the model’s outputs change \cite{tedre2021ct}. However, evidence remains limited on how students and teachers develop accurate mental models of data-driven systems and the specific concepts and skills that support that development. Strengthening this line of research likely requires interdisciplinary grounding, particularly from mathematics and statistics education \cite{Shapiro2018}. We propose two dimensions we argue are central to incorporating data science activities into K--12 computing education.

\subsection{Knowledge-based vs data-driven models}
\label{ref:KBvsDD}
In CS education, students typically solve problems through algorithmic solutions (e.g., programming). This approach, termed \textit{rule-based} (or \textit{knowledge-based} or \textit{symbolic}), is based on the idea of \textit{``logical proof of correctness''} \cite[p.~27]{Shapiro2018} and aligns with a traditional algorithmic view of problem solving in which a solution is formalised and implemented \cite{Tedre2020,tedre2021ct}. This contrasts with a data-driven approach where students instead train and evaluate models using data and evaluate systems through a \textit{``statistical demonstration of effectiveness''} \cite[p.~27]{Shapiro2018}. \citeauthor{Shapiro2018} describes these as two notional machines, namely the \textit{``classical logical computer''} and the \textit{``statistical model''} \cite[p.~28]{Shapiro2018}.

\subsection{Explainability of models}
\label{ref:highLowExplainability}
In many ML systems and tools, certain processes are often abstracted away (i.e., \textit{black-boxed}) so either the inner workings of the learning algorithm and/or the role of data in shaping model outcomes may be less visible to learners \cite{MoralesNavarroKafai2024}. A pedagogical challenge for educators, then, is the extent to which data-driven models can be explained \cite{Kim2024XAI}. Explainability varies across models, including how transparent, interpretable, and understandable a model’s internal logic is to end users \cite{Budhkar2025Demystifying}. Explainability can decrease as dimensionality increases since humans cannot readily inspect or reason over all components of a high-dimensional model at once \cite{flora2022xai}. For example, a low-dimensional linear regression can be considered highly explainable because its decisions can be accounted for in terms of explicit parameters and relationships \cite{flora2022xai}. By contrast, data-driven models such as recurrent neural networks (RNNs) or random forests are typically harder to explain because the pathways by which they produce decisions are comparatively opaque \cite{Budhkar2025Demystifying}. Strengthening the explainability of such models, especially where decision-making processes are not transparent, is frequently presented as important for establishing trust and supporting reliable use \cite{LiptonModelInterpret}.


Within explainable AI (XAI), \textit{post-hoc} techniques aim to provide explanations of otherwise opaque models \cite{flora2022xai}. Common examples include \textit{feature importance} methods that indicate which inputs most influenced a particular output, and counterfactual explanations, which illustrate the smallest changes needed to obtain a different prediction \cite{Kim2024XAI}. By contrast, \textit{ante-hoc} explanations, sometimes termed ``intrinsic explainability''~\cite[p.~2]{RetzlaffxAI}, are associated with models whose structure is itself transparent, such as \textit{``linear regression, decision tree models, k-nearest neighbors''} \cite[p.~348]{Budhkar2025Demystifying}, where input–output relationships are more directly legible. From an educational perspective, this introduces a tension: if learners’ early experiences with rule-based programming formulated an expectation that computational systems are transparent and verifiable, they may struggle to reconcile that expectation with the limited transparency often characteristic of more opaque data-driven systems \cite[p.~28]{Shapiro2018}.


\subsection{Summary}
\label{ref:literatureSummary}
As CS education evolves to incorporate data literacy skills, questions remain over \textit{what} and \textit{how} to teach, particularly given the interdisciplinary nature of data science. To address this \textit{``fractured''} issue \cite[p.~248]{CruickshankWholeElephant}, we focus on two dimensions that could provide a common vocabulary to describe learning environments across an emerging landscape: (i) logic (i.e., knowledge-based versus data-driven), and (ii) explainability (i.e., transparent versus opaque models). We review K--12 interventions through these dimensions to characterise existing data literacy experiences and identify shared data literacy pathways.

\section{Method}
\label{sect:method}
\subsection{Systematic literature review}
\label{sect:slr}
We conducted a systematic review of literature relating to the teaching and learning of data literacy skills in K--12 settings \cite{Whyte2026SLR}. Further details on the screening process and criteria, as well as a copy of the literature review data, are available on the study website for further reading \cite{datasciencestudywebpage}. Following the PRISMA 2020 guidelines \cite{Page2021PRISMA}, we sourced peer-reviewed empirical studies that (i) related to data literacy in K--12 settings; (ii) were published between 2019 and 2024; (iii) were published in computing education and other relevant fields (e.g., in mathematics education, STEM fields, learning sciences) as data literacy is interdisciplinary in nature; and (iv) took place in classroom-based activities and non-formal contexts (e.g., data camps).

\subsection{Data analysis}
\label{sect:dataAnalysis}
\textbf{Data paradigms:} We analysed learning activities against two dimensions: (i) whether students engaged in either creating knowledge-based or data-driven models \cite{Olari2024Concepts} and (ii) whether these models were transparent or opaque \cite{flora2022xai,Budhkar2025Demystifying}. Comparing these two dimensions against one another led to four distinct quadrants (or \textit{data paradigms}) in which student activities took place (see the data paradigms framework in Table \ref{tab:framework}). We were also interested to source interdisciplinary literature (computing education, statistics and mathematics, and domain-specific applications) and consider how terminology and language differ across disciplinary boundaries. 

\begin{table*}[htb]
\caption{Data paradigms framework}
\label{tab:framework}
\centering
\begin{tabular}{|p{3cm}|p{5cm}|p{5cm}|} \hline
\textbf{\textit{Dimension}} & \textbf{Knowledge-based (KB)} & {\textbf{Data-driven (DD)}} \\ \hline
\textbf{Transparent (T)}  \textit{Explainable by design} & \textit{KB+T}: Rule-based models which are explainable by design (e.g., rule-based decision trees, manual classification) & \textit{DD+T:} Data-driven models which are explainable by design (e.g., linear regression, k-nearest neighbour) \\ \hline
\textbf{Opaque (O)} \textit{Explainable through additional methods (e.g. post-hoc)} & \textit{KB+O:} Rule-based models which are explainable through additional methods & \textit{DD+O:} Data-driven models (e.g., neural networks, random forest) which are explainable through additional (e.g. post-hoc) methods \\ \hline
\end{tabular}
\end{table*}

Similar to the quadrant model \cite{KlahrQuadrant}, we ascribed learning activities to each paradigm (or `quadrant') based on students' engagement with knowledge-based or data-driven systems, and the extent to which these were explainable, within learning activities. With the advancement of post-hoc explanation methods, we recognise that this classification may evolve over time and activities could be reclassified. 

Interventions were generally designed for secondary-aged students (12--18 years old) (\textit{n}=191, 66.8\% of instances) with fewer reported for primary-aged students or younger (3--11 year-olds) (\textit{n}=89, 31.1\%); several studies provided no age data (\textit{n}=6). Fewer examples of interventions were designed for 18-year-old students though this is likely to reflect when tertiary education begins in most country contexts. 

\textbf{Data literacy trajectories:} As some papers featured multiple activities that were classified across multiple paradigms, we wanted to consider how learners progressed from knowledge-based to data-driven, and from transparent to more opaque, systems. We therefore visualised activities across the four quadrants proposed in our data paradigms framework \cite{Whyte2026SLR} to consider whether and how interventions moved between each paradigm. Drawing on the notion of learning trajectories in mathematics education \cite{ClementsHLT} and computing education \cite{RichLearningTrajectories}, we inductively determined what current trajectories---in terms of the underlying learning goals---are employed within the analysed interventions.

Three authors qualitatively coded the 84 papers until consensus was reached on the final coding scheme and two authors coded the remaining papers. Though some activities were more challenging to categorize, and required discussion to reach consensus, we nonetheless calculated inter-rater reliability with a Cohen’s Kappa value of $\kappa=0.85$, indicating strong agreement \cite{cohenkappainterrater}. This process resulted in four distinct learning trajectories (see Figures \ref{fig:transparent}, \ref{fig:datadriven}, \ref{fig:jumping}, and \ref{fig:bridging}) that characterised student learning within interventions. These are described in detail in the following section.

\section{Findings}
\label{sect:findings}
Most studies were conducted in the United States (\textit{n}=46, 54.8\%), followed by Germany (\textit{n}=6, 7.1\%) and Finland (\textit{n}=5, 6.0\%). Australia, Brazil, Denmark, Hong Kong, Israel, Japan, Nigeria, South Korea, and Taiwan each contributed two studies (\textit{n}=2, 2.4\% each), and examples from Austria, China, Colombia, India, Singapore, Thailand, the Netherlands, Spain, and Switzerland accounted for one study each (\textit{n}=1, 1.2\% each). In the following sections, we next outline some emerging trajectories that were identified across the reviewed literature.

\subsection{Single paradigm activities}
\label{sect:singleParadigm}
Most of the reviewed studies describe learning experiences that remain within a single paradigm (\textit{n}=57; \textbf{KB-T}=1; \textbf{DD-T}=36; \textbf{DD-O}=20) \cite{Whyte2026SLR}. Activities within this quadrant include discipline-specific activities, such as modelling in science \cite{GuyGaytan2019} and data analysis using R in mathematics \cite{Wiedemann2020}. Many examples focus on `inspectable' data practices such as visualising and interpreting datasets within the data-driven/transparent paradigm (\textbf{DD-T}) \cite{ArastoopourIrgens2024}, whereas others centre on data-driven/opaque (\textbf{DD-O}) activities, such as image classification in computing education, where model reasoning is not made directly inspectable \cite{Tseng2024CoML}. Single-paradigm designs also appear in discipline-specific implementations, where the activity is tightly aligned with subject-specific goals and tools (e.g., data analysis using R in mathematics \cite{Wiedemann2020}), which can prioritise domain learning outcomes over cross-paradigm comparison or progression.

\subsection{Trajectory \#1: Keeping transparent}
\label{sect:Transparent}
\begin{figure}[htbp]
  \centering
  \begin{minipage}{0.48\textwidth}
    \centering
    \includegraphics[width=\textwidth]{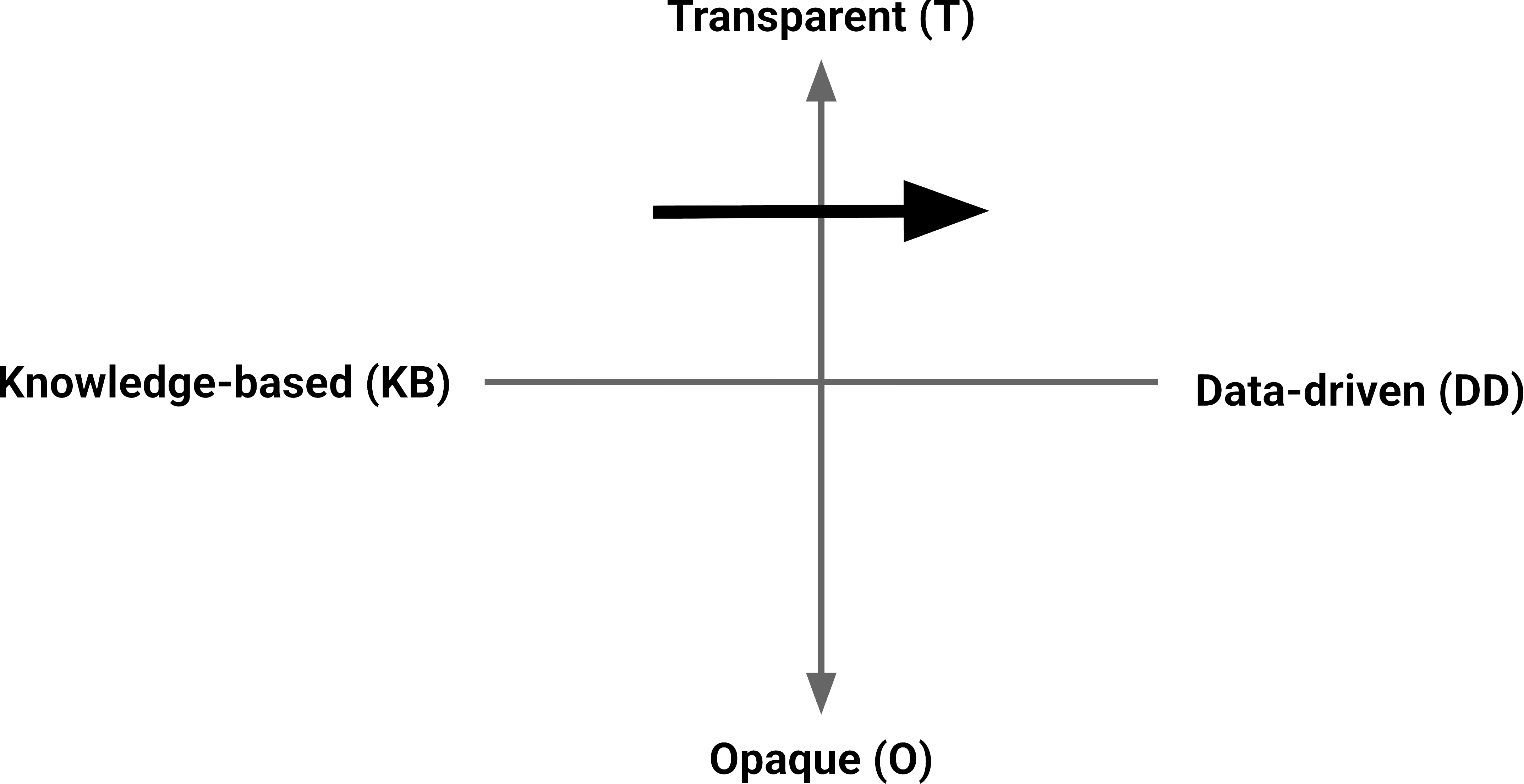}
    \caption{Trajectory \#1: Keeping transparent}
    \label{fig:transparent}
  \end{minipage}
  \hfill 
  \begin{minipage}{0.48\textwidth}
    \centering
    \includegraphics[width=\textwidth]{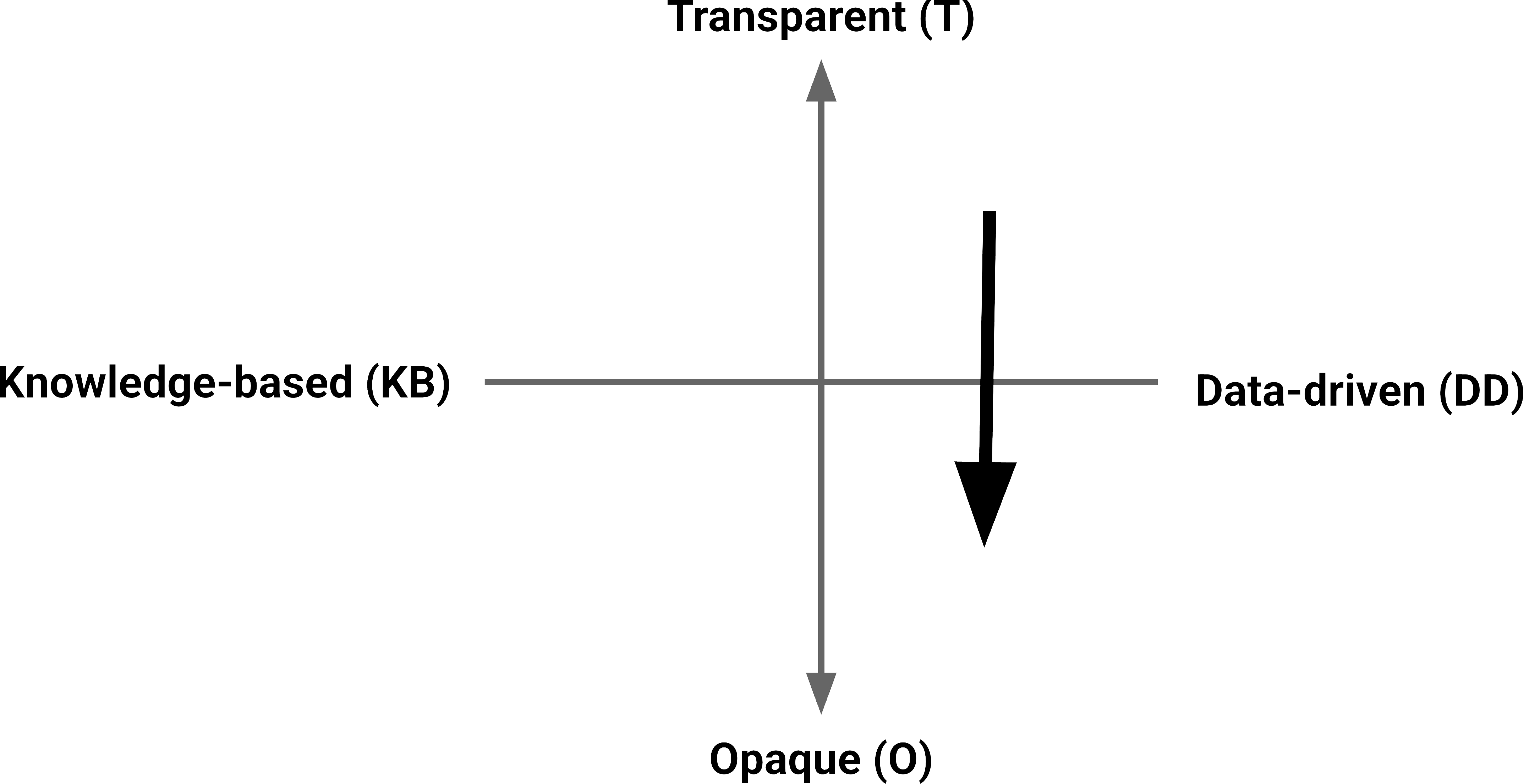}
    \caption{Trajectory \#2: Keeping data-driven}
    \label{fig:datadriven}
  \end{minipage}
\end{figure}

In this trajectory, learners move from the knowledge-based/transparent (\textbf{KB-T}) to the data-driven/transparent (\textbf{DD-T}) paradigm (see Figure \ref{fig:transparent}). 11 of the 84 studies followed this pathway in the SLR \cite{Whyte2026SLR} (\textit{n}=11). For example, \citeauthor{Jiang2022} uses \textit{StoryQ}, a web-based ML and text mining tool for young learners to move from identifying sentiment cues in text (e.g dessert reviews), to refining features to improve classification model accuracy through error analysis \cite{Jiang2022}. Similarly, \citeauthor{Kajiwara2023-ed} justify decision trees as \textit{``white-box machine learning''}, in contrast to \textit{``black-box''} approaches such as K-Nearest Neighbor (KNN) and neural networks, positioning rule-based `transparency' as a scaffold for an interpretable data-driven classifier \cite[p.~4]{Kajiwara2023-ed}. 

\subsection{Trajectory \#2: Keeping data-driven}
\label{sect:dataDriven}
In this trajectory, learners move from data-driven/transparent (\textbf{DD-T}) to data-driven/opaque (\textbf{DD-O}) (see Figure \ref{fig:datadriven}) (\textit{n}=5). Students typically begin with exploratory analysis or simple supervised models (e.g., scatter plots \cite{Kim2024-fo}, regression/classification prior to CNNs \cite{Kong2023}, linear regression on self-collected temperature data\cite{Lu2023-gs}) and then progress to less transparent classifiers or deep learning for prediction or recognition (e.g., K-Nearest Neighbor (KNN) and/or Support Vector Machine (SVM) on historical weather data \cite{Lu2023-gs} or KNN followed by image recognition based on deep learning \cite{Fujishima}). In \citeauthor{Lin2020}, the conversational agent \textit{Zhorai} shifts from an inspectable, user-data-driven construction to a less transparent classification step summarised via a histogram of word-similarity scores \cite{Lin2020}.

\subsection{Trajectory \#3: Jumping}
\label{sect:multipleParadigms} 
\begin{figure}[htbp]
  \centering
  \begin{minipage}{0.48\textwidth}
    \centering
    \includegraphics[width=\textwidth]{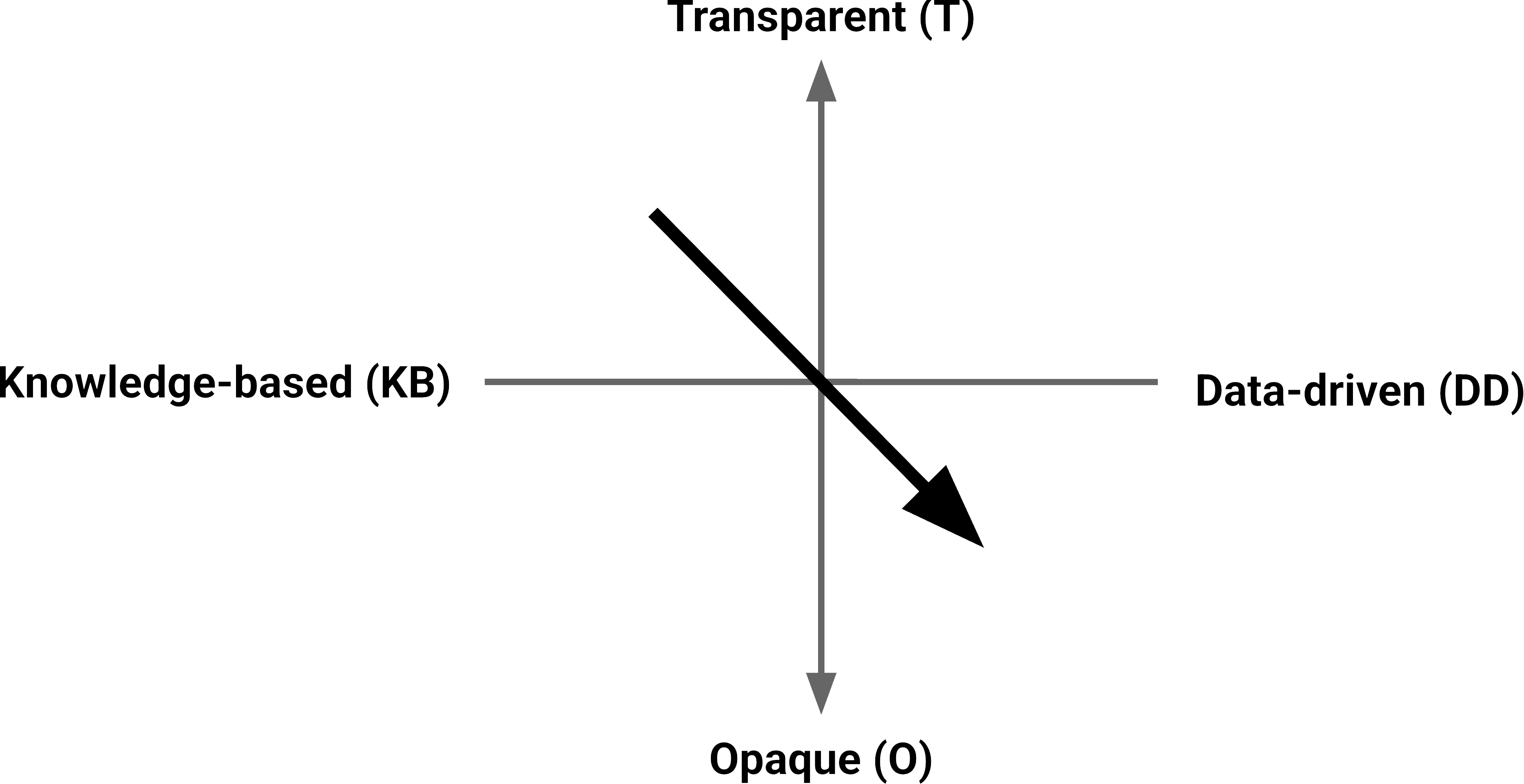}
    \caption{Trajectory \#3: Jumping}
    \label{fig:jumping}
  \end{minipage}
  \hfill 
  \begin{minipage}{0.48\textwidth}
    \centering
    \includegraphics[width=\textwidth]{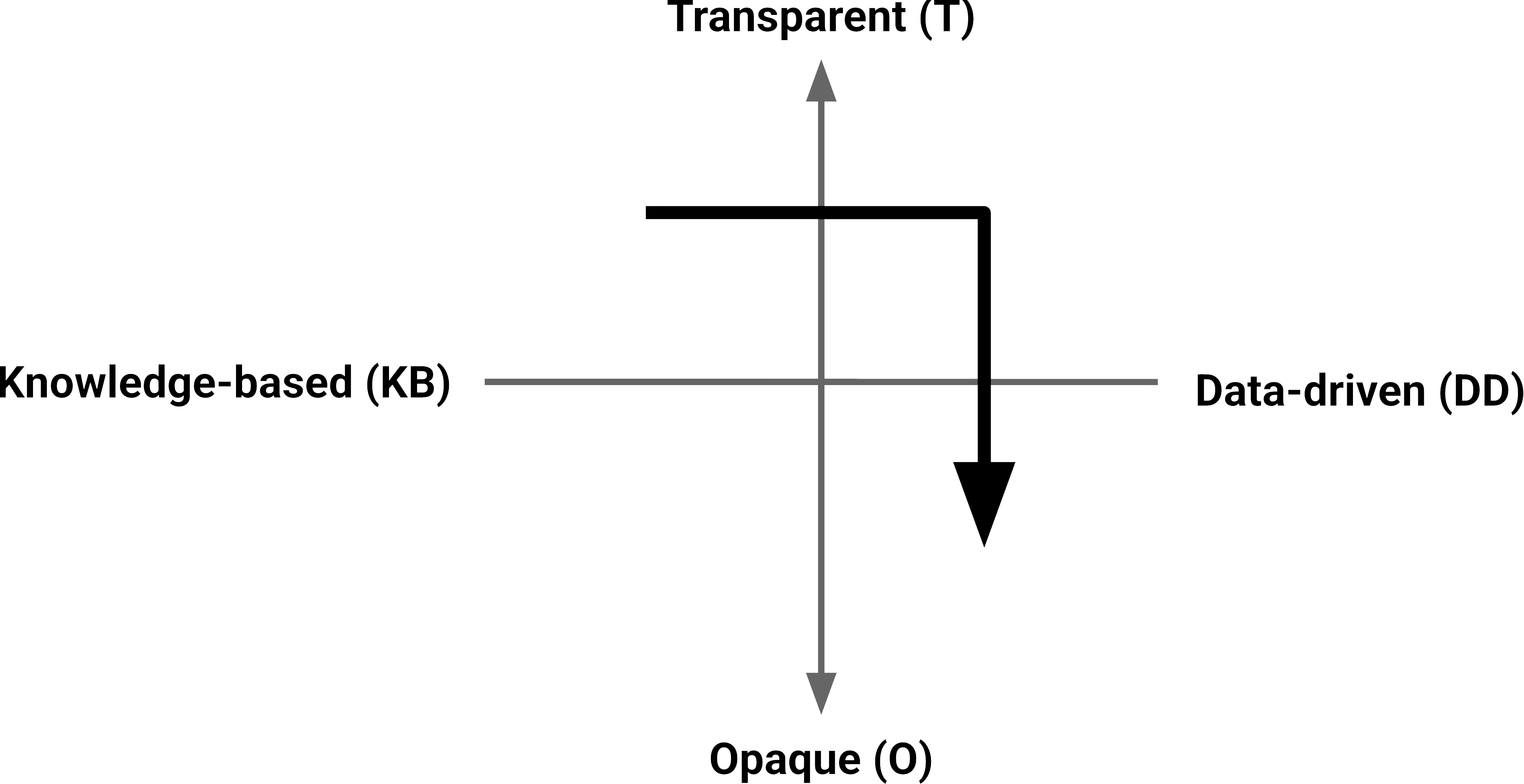}
    \caption{Trajectory \#4: Bridging}
    \label{fig:bridging}
  \end{minipage}
\end{figure}

In this trajectory, learners jump from the knowledge-based/transparent (\textbf{KB-T}) to the data-driven/opaque (\textbf{DD-O}) paradigm (see Figure \ref{fig:jumping}) (\textit{n}=9) \cite{Famaye2025, Guerreiro-Santalla2022-nj, ArastoopourIrgens2022, Shamir2022, Simbeck2024, Van_Brummelen2020, Williams2019-pj, Williams2022, Zhang2023}. One activity moved from \textit{`teaching computers through programming'} (i.e. creating rules in Scratch; \textbf{KB-T}) to \textit{``teaching computers to learn''} \cite[p. 377]{Simbeck2024} (i.e. machine learning; \textbf{DD-O}). Another distinguished between traditional CT practices (e.g., decomposition) and \textit{``ML CT practices''} \cite[p. 4]{Shamir2022} (e.g., feature selection). Students construct a \textit{``rule-driven ML system''} consisting of logic gates and truth tables (\textbf{KB-T}), whereas in the \textit{Learning ML by teaching} course, students create a \textit{``data-driven ML system''} using \textit{Machine Learning for Kids} for classification (\textbf{DD-O}). \citeauthor{ArastoopourIrgens2022} discussed their approach to ``Critical Machine Learning'' education program where one activity moved from human-created rules (Pizza algorithm) to thinking critically about and interacting with data-driven systems through \textit{QuickDraw!} and \textit{Teachable Machines} by discussing algorithm's performance limitations, specifically its inability to recognise every drawing \cite{ArastoopourIrgens2022}. In another example, students take part in a workshop based on ML-based conversational AI, where simple rule-based conversational agents are introduced to \textit{``provide a segue''} into developing more complex (and increasingly opaque) ML-based agents \cite{Van_Brummelen2020}. Across these studies, the jump is typically framed as a shift from explicit, human-authored rules and `classical' CT to train systems whose behaviour must be inferred from data rather than inspected directly.

\subsection{Trajectory \#4: Bridging}
\label{sect:bridging}
Moving students through more than two paradigms (Figure \ref{fig:bridging}) was not common with only two studies exemplifying this \cite{Whyte2026SLR}. A shared pattern is that opaque ML work is not introduced as a conceptual jump, as in the \textit{Jumping} trajectory, but is sequenced from transparent reasoning and/or inspectable data work. In \citeauthor{BrollGrover2024}, learners begin with rule-based explanations through Denial-of-Service (DoS) attacks (\textbf{KB-T}), then explore a Twitter dataset using CODAP to visualise the data and provide ideas for classification (\textbf{DD-T}), before training a simple form of generative adversarial network \cite[p. 15995]{BrollGrover2024} (\textbf{DD-O}) to explore ideas of the generator and discriminator. In another example, \citeauthor{Napierala2023} introduce a curriculum for supervised ML, which transitions students through different paradigmatic concepts. Initially, students engage with a set of leaf image training data (\textbf{DD-T}) to discover features, which is followed by the manual creation of a decision tree (\textbf{KB-T}) to develop explicit classification rules. Students then reflect on the principles and limitations of \textit{Seek}, a leaf identification application (\textbf{DD-O}), completing a journey that bridges transparent, rule-based reasoning with data-driven, opaque activities \cite{Napierala2023}. These curricula are structured so that learning experiences \textit{bridge} between transparent and opaque, through a transparent data-driven activity, drawing attention to how concepts differ across the paradigms.

\section{Discussion}
\label{sect:discussion}
Researchers have argued that data-driven technologies (e.g., ML) necessitate a need for data literacy skills \cite{tedre2021ct,LongMagerko}. In order to critically engage with these technologies, educators and resource need to consider what emerging data literacy skills are needed \cite{Olari2024Practices}. In CS education, we are particularly interested in exploring how we transition educators and learners from traditional knowledge-based (or rule-based) logic towards data-driven (e.g. ML) reasoning. To that end, we have articulated how the data paradigms framework \cite{Whyte2026SLR} and associated trajectories could provide a starting point for researchers, educators and resource developers to compare how data literacy learning experiences position learners with respect to knowledge-based versus data-driven reasoning and transparent versus opaque systems.

\subsection{Articulating a shared language for data literacy goals}
\label{sect:language}
One enduring challenge, particularly within CS education, is the lack of a shared vocabulary when discussing the nature of data literacy and deciding on common goals. In our examples, we found discrepancies in how approaches are discussed, such as transparent knowledge-based approaches (\textbf{KB+T}) defined as \textit{``rule-driven learning''} \cite[p. 2]{Shamir2022}, \textit{``teaching computers through programming''} \cite[p. 375]{Simbeck2024} or even \textit{``rule-based AI''}~\cite[p. 15657]{Van_Brummelen2020}. Conversely, opaque data-driven approaches (\textbf{DD+O}) are referred to as \textit{``ML-based data-driven thinking''} \cite[p. 1]{Shamir2022}, \textit{``ML-based AI''}~\cite[p. 15657]{Van_Brummelen2020} or even \textit{``CT 2.0''}~\cite{tedre2021ct}. 

These diverse conceptualizations highlight the need for shared vocabularies and evaluative approaches that travel across adjacent literacies and disciplinary contexts \cite{CHI_EA_2026_DataLiteracy_Workshop}. It also underscores concerns that data science is often fragmented across disciplinary traditions \cite{CruickshankWholeElephant}. Making paradigms and trajectories explicit may therefore function as a translation device for comparing interventions across diverse disciplines and literature. We recognize that our proposed trajectories are just that---\textit{propositions}---and welcome further discussion on how these should be defined, what language is useful in characterizing the distinction between paradigms, and how learning differs across trajectories.

\subsection{Mapping common data literacy trajectories}
\label{sect:clearTrajectories}
In this position paper, we suggest that the two proposed dimensions---logic (knowledge-based vs data-driven) \cite{tedre2021ct} and explainability (transparent vs opaque) \cite{Budhkar2025Demystifying}---could serve to delineate the epistemic boundaries of diverse data literacy activities. We have also argued that attention be given to how what approaches to \textit{scaffolding} \cite{TabakKyza2018} are needed to move learners between these paradigms. Different trajectories may be appropriate depending on instructional aims and the kinds of use cases students need. Rather than prescribing a trajectory, we propose that these dimensions make visible \textit{what} should be taught explicitly for a given paradigm and trajectory to support thorough understanding. For example, introducing opaque models requires more explicit attention to concepts such as model confidence \cite{ArastoopourIrgens2022,BrollGrover2024} and evaluating model performance using confusion matrices \cite{Jiang2022,Kim2024-fo}. Conversely, introducing transparent models requires a greater focus on practices such as feature weighting \cite{Jiang2022} or visualising relationships between variables using a heatmap of correlation coefficients \cite{Kim2024-fo}. Exposing students to more `transparent' (i.e. glass-box) processes has been argued as necessary to support later engagement with more `opaque' systems \cite{HitronBlackBox}.

Few examples within the \textit{Bridging} trajectory were found which suggests that the distinction between rule-based and data-driven paradigms is not commonly taught \cite[e.g., ][]{tedre2021ct,BrollGrover2024, Napierala2023}. Activities that purposefully move learners between multiple paradigms---including those that revisit earlier concepts and skills---might support a deeper understanding of the epistemic differences across paradigms. For example, the \textit{Keeping transparent} trajectory \cite{Kajiwara2023-ed,Jiang2022} seemingly leverages learners’ prior expectations that rule-based systems are inspectable and predictable while gradually shifting from hand-authored rules to data-driven models while preserving explainability.

Likewise, the \textit{Keeping data-driven} trajectory foregrounds data-driven reasoning while decreasing the level of transparency. These approaches may require additional scaffolding as learners' ability to understand how opaque systems work diminishes.

In the \textit{Jumping} trajectory, learners `jump' from handling rule-based systems to engaging with opaque data-driven models, including neural networks. Without an understanding of how data-driven models work (e.g., through explainable tools and approaches), learners may have incorrect assumptions about the level of `correctness' of data-driven systems and may ascribe certainty to their outputs. 

We present the framework and associated trajectories to provoke debate on their usefulness for mapping the progression of learning across data paradigms. As we have not advocated for any particular trajectory as `optimal', especially when moving from teaching knowledge-based to data-driven systems, we are interested in how to define these trajectories, and which will serve educators most effectively. By being explicit and juxtaposing differences in reasoning across paradigms, students can understand the benefits and trade-offs associated with various levels of explainability as well as the underlying logic (i.e. data-driven vs rule-based). Though few in number, these instances were encouraging though they point to specific curricular and/or pedagogical challenges, including context demands, a lack of suitable tools, and the technical complexity as learners and educators move toward less transparent systems. 

\section{Limitations and future work}
\label{sect:limitationsFutureWork}
Our results are limited to the extent that we analysed early examples of research-led interventions rather than curricula or grey literature. As such, results are likely to have fewer examples of progressions of learning than more lengthy materials. To that end, we found that most interventions remained within a single paradigm (\textit{n}=57/84), most often in \textbf{DD-T} or \textbf{DD-O} activities (see section~\ref{sect:singleParadigm})---such as classification with ML tools \cite{Theofanos2024AI}---with few examples of instructional sequences that moved between these paradigms.

Likewise, our understanding of what data literacy skills are needed in CS education is still limited. As curricula are being readily developed to support data literacy, future work could apply the data paradigms framework to curricula and associated learning materials to identify further trajectories and progressions between different paradigms. Further, research could focus on articulating and enacting different learning sequences and using the data paradigms framework to map their trajectories.

\section{Conclusion}
\label{sect:conclusion}
In this position paper, we present the data paradigms framework and propose a set of learning trajectories as a tool to map the landscape of K–12 data literacy initiatives. We suggest these tools could provide a starting point for researchers and resource developers when considering the design of data literacy learning environments. Our findings demonstrate how some environments successfully scaffold learners from transparent rules to transparent data models (\textit{Keeping Transparent}). However, we found few instances where all three critical domains are touched upon (or the \textit{Bridging} trajectory): knowledge-based logic, then transparent (or inspectable) data analysis, before encountering opaque ML systems. The \textit{Jumping} trajectory also indicates a pedagogical challenge as learners `bypass' critical skills in understanding the influence of data on model behaviour. We argue that these skills are necessary to prepare students when encountering `black-boxed' systems in later learning experiences.

Beyond CS education, we argue the data paradigms framework and associated trajectories may offer a shared vocabulary for learning scientists and CS researchers to map the epistemic shifts encountered by learners as they move across contexts and disciplines. We hope to further discuss this work as part of  and consider how we might align data literacy goals across disciplines and better prepare students for a data-driven future.


\bibliographystyle{ACM-Reference-Format}
\footnotesize
\bibliography{datascienceslr}

\end{document}